\documentclass[runningheads]{llncs}
\usepackage[T1]{fontenc}
\usepackage{graphicx}
\usepackage{cite}
\usepackage{amsmath,amssymb,amsfonts}
\usepackage{graphicx}
\usepackage{textcomp}
\usepackage{algorithm,algorithmic}
\usepackage{multicol}
\usepackage{multirow}
\usepackage{changepage,threeparttable}
\usepackage{makecell}
\usepackage{array}
    \newcolumntype{P}[1]{>{\centering\arraybackslash}p{#1}}
    \newcolumntype{M}[1]{>{\centering\arraybackslash}m{#1}}
\usepackage{booktabs, multirow}
\usepackage{soul}
\usepackage[T1]{fontenc}

\usepackage{amsmath,amsfonts}
\usepackage{algorithmic}
\usepackage{array}
\usepackage[caption=false,font=normalsize,labelfont=sf,textfont=sf]{subfig}
\usepackage{textcomp}
\usepackage{stfloats}
\usepackage{url}
\usepackage{verbatim}
\usepackage{graphicx}
\def\BibTeX{{\rm B\kern-.05em{\sc i\kern-.025em b}\kern-.08em
    T\kern-.1667em\lower.7ex\hbox{E}\kern-.125emX}}
\usepackage{balance}

\usepackage[colorlinks,urlcolor=blue,linkcolor=blue,citecolor=blue]{hyperref}

\usepackage{color,array}
\usepackage[utf8]{inputenc}
\usepackage{amsmath,amssymb,amsfonts}
\usepackage{algorithm,algorithmic}
\usepackage[mathscr]{euscript}

\usepackage{orcidlink}

\begin{document}
\title{GENRE-CMR: Generalizable Deep Learning for Diverse Multi-Domain Cardiac MRI Reconstruction}
%
%


\author{Kian Anvari Hamedani\inst{1}\orcidlink{0009-0005-7416-1987} \and
Narges Razizadeh\inst{1}\orcidlink{0009-0007-8985-9653} \and
Shahabedin Nabavi\inst{1,*}\orcidlink{0000-0001-7240-0239} \and
Mohsen Ebrahimi Moghaddam\inst{1}\orcidlink{0000-0002-7391-508X}}
\authorrunning{K. Anvari Hamedani et al.}
\institute{
Faculty of Computer Science and Engineering, Shahid Beheshti University, Tehran, Iran \\
\email{\{k.anvarihamedani,n.razizade\}@mail.sbu.ac.ir} \\
\email{\{s\_nabavi,m\_moghadam\}@sbu.ac.ir}\\
\href{https://github.com/kiananvari/GENRE-CMR}{https://github.com/kiananvari/GENRE-CMR}
}
\maketitle              

\begin{abstract}
Accelerated Cardiovascular Magnetic Resonance (CMR) image reconstruction remains a critical challenge due to the trade-off between scan time and image quality, particularly when generalizing across diverse acquisition settings. We propose GENRE-CMR, a generative adversarial network (GAN)-based architecture employing a residual deep unrolled reconstruction framework to enhance reconstruction fidelity and generalization. The architecture unrolls iterative optimization into a cascade of convolutional subnetworks, enriched with residual connections to enable progressive feature propagation from shallow to deeper stages. To further improve performance, we integrate two loss functions: (1) an Edge-Aware Region (EAR) loss, which guides the network to focus on structurally informative regions and helps prevent common reconstruction blurriness; and (2) a Statistical Distribution Alignment (SDA) loss, which regularizes the feature space across diverse data distributions via a symmetric KL divergence formulation. Extensive experiments confirm that GENRE-CMR surpasses state-of-the-art methods on training and unseen data, achieving 0.9552 SSIM and 38.90 dB PSNR on unseen distributions across various acceleration factors and sampling trajectories. Ablation studies confirm the contribution of each proposed component to reconstruction quality and generalization. Our framework presents a unified and robust solution for high-quality CMR reconstruction, paving the way for clinically adaptable deployment across heterogeneous acquisition protocols.

\keywords{Accelerated MRI \and Cardiac MRI \and CMR Reconstruction.}
\end{abstract}
\section{Introduction}
Cardiovascular Magnetic Resonance (CMR) imaging plays a pivotal role in non-invasive cardiovascular assessment, offering high-resolution, multiparametric information without ionising radiation \cite{arnold2020cardiovascular}. It is considered the gold standard for evaluating cardiac function, perfusion, viability, fibrosis, and congenital abnormalities \cite{vasquez2019clinical}. However, widespread clinical adoption remains constrained by the inherently slow acquisition process, driven by sequential k-space sampling and the need for high spatial and temporal resolution. This makes CMR sensitive to physiological motion (cardiac, respiratory), which can introduce artifacts and compromise image quality \cite{nabavi2023generalised}. Recent studies have also investigated deformable registration and motion estimation strategies to better capture spatiotemporal cardiac dynamics \cite{zakeri2023dragnet, bi2024segmorph, kebriti2025fractmorph}. Additionally, the complex cardiac anatomy and diverse imaging protocols, spanning various contrasts (Cine, T1/T2 Mapping, LGE), trajectories (e.g., uniform, 3D k-t Gaussian/Radial), and anatomical views (e.g., long-axis, short-axis, aortic), prolong scan time and increase reconstruction complexity \cite{enders2011reduction, wang2025cmrxrecon2024}. Differences in scanner hardware and software further hinder robust and consistent reconstruction across sites \cite{guan2021domain}.

Recent advances in deep learning (DL) have shown promise in accelerating CMR image reconstruction by learning direct mappings from undersampled k-space to high-quality images \cite{anvari2024all, xin2024rethinking, xu2024hypercmr, xu2024segmentation, yiasemis2024deep}. Leading solutions from the CMRxRecon challenges \cite{wang2024cmrxrecon, wang2025cmrxrecon2024} have achieved superior performance over traditional methods \cite{pruessmann1999sense, griswold2002generalized, lustig2007sparse}, as reviewed in \cite{lyu2025state, wang2025towards}. However, DL models often fail to generalize well in clinical settings due to domain shifts arising from variability in acquisition protocols, scanner vendors, anatomical coverage, and patient characteristics \cite{guan2021domain}. Addressing this generalization gap, through robust domain adaptation and domain-robust reconstruction strategies, is critical for safe and reliable deployment of DL-based reconstructions in real-world clinical practice \cite{nabavi2024multiple, nabavi2024statistical}.

Several recent works have explored domain-robust MRI reconstruction. One approach \cite{ouyang2019generalizing} demonstrated that networks trained with natural image datasets containing synthesized phase information can generalize across unseen contrasts and anatomies. A self-supervised method \cite{millard2024clean} introduced Robust SSDU, which is capable of recovering clean reconstructions even from noisy and sub-sampled training data. Another framework \cite{gao2024mrpd} proposed MRPD, which leverages large latent diffusion models pre-trained on natural images and adapts them to MRI reconstruction, demonstrating unprecedented cross-domain generalizability in unsupervised settings. More recently, LowRank-CGNet \cite{patel2024low} was presented, integrating low-rank tensor modeling with conjugate gradient data consistency to handle diverse anatomy, contrast, and undersampling artifacts. While these approaches provide important advances, many rely on heavy diffusion models, external natural image priors, or structural assumptions, and do not jointly integrate unrolled optimization, adversarial learning, and explicit domain alignment.

To address these challenges, we propose a generalizable CMR reconstruction framework based on a generative adversarial network (GAN) architecture. The generator incorporates a residual deep unrolled network that mimics compressed sensing-based iterative optimization, progressively refining intermediate k-space estimates from undersampled input data. To improve anatomical fidelity and reduce blurring, we introduce Edge-Aware Reconstruction (EAR) loss that emphasizes recovery of clinically relevant boundaries. To address domain variability, we include a Statistical Distribution Alignment (SDA) loss that aligns latent features across different CMR domains. Furthermore, inspired by recent work in prompt-based MRI reconstruction \cite{xin2023fill}, we integrate prompt learning to enable adaptive reconstruction across diverse contrasts, trajectories, and anatomical views within a unified model. Our main contributions are:

\begin{itemize}
    \item We present a generalizable residual deep unrolled reconstruction framework for CMR, which integrates compressed sensing-based inverse problem solving into the GAN generator. This design enables accurate reconstruction from highly undersampled k-space data. Experiments on the CMRxRecon 2025 dataset demonstrate that our method outperforms existing state-of-the-art approaches in both image quality and generalization across out-of-distribution scenarios.
    \item We propose EAR loss, which enhances the reconstruction of fine anatomical details and mitigates the blurring artifacts common in deep learning-based methods.
    \item To address the challenge of domain shift caused by variability in contrasts, trajectories, anatomical coverage, and scanner settings, we incorporate SDA loss that explicitly reduces distributional discrepancies across different CMR domains, promoting robust generalization.
\end{itemize}

\section{Materials and Methods}

\subsection{Dataset}
We evaluated our method on the CMRxRecon 2025 dataset \cite{b6xs-gv29-25}, a large scale, multi-center, multi-vendor benchmark specifically designed to assess the robustness of cardiac MRI reconstruction models across clinically diverse settings. The dataset comprises over 600 subjects collected from multiple institutions and scanner vendors, including GE, Philips, Siemens, and UIH. It covers a broad patient population, including healthy volunteers, individuals with various cardiac pathologies such as cardiomyopathies, myocardial infarction, and arrhythmias, as well as pediatric cases. It features a wide range of CMR modalities (e.g., Cine, T1/T2 Mapping, LGE, and perfusion), along with varying sampling trajectories (Cartesian, Radial, and Gaussian) and magnetic field strengths (1.5T, 3T, and 5T). Each case includes fully sampled and under sampled k-space data, sampling masks, and ground truth reconstructions.

\subsection{A Residual Deep Unrolled Reconstruction Framework}
Building on our 2024 all-in-one Patch-GAN reconstruction model \cite{anvari2024all}, this framework introduces residual cascaded unrolling with improved connectivity, while augmenting the original loss formulation with the proposed EAR and SDA terms. Specifically, we extend the earlier unrolled model by introducing residual connections between consecutive cascaded modules, enabling more effective feature propagation and mitigating vanishing gradients during training. The architecture consists of a generator and an unrolled discriminator trained adversarially. The generator follows a cascaded structure that unrolls an iterative reconstruction process, progressively refining the under-sampled k-space at each step. These residual paths directly pass feature maps from shallow reconstructors to deeper ones, encouraging the reuse of early-stage representations and promoting better convergence and reconstruction accuracy. The model accepts multi-coil k-space data from various vendors (Philips, Siemens, UIH, and GE), and different acquisition protocols, accounting for real-world domain shifts across centers. The network learns a sensitivity map and uses coil-combination layers to generate intermediate and final reconstructions. In addition, an unrolled discriminator is employed to enhance the reconstruction quality through adversarial training.

The complete procedure for our proposed GENRE-CMR framework is formalized in Algorithm \ref{alg:genre}, which integrates the residual unrolling, adversarial training, and the proposed EAR and SDA losses into a unified reconstruction pipeline.

\begin{figure}[!t]
\centerline{\includegraphics[width=1\textwidth]{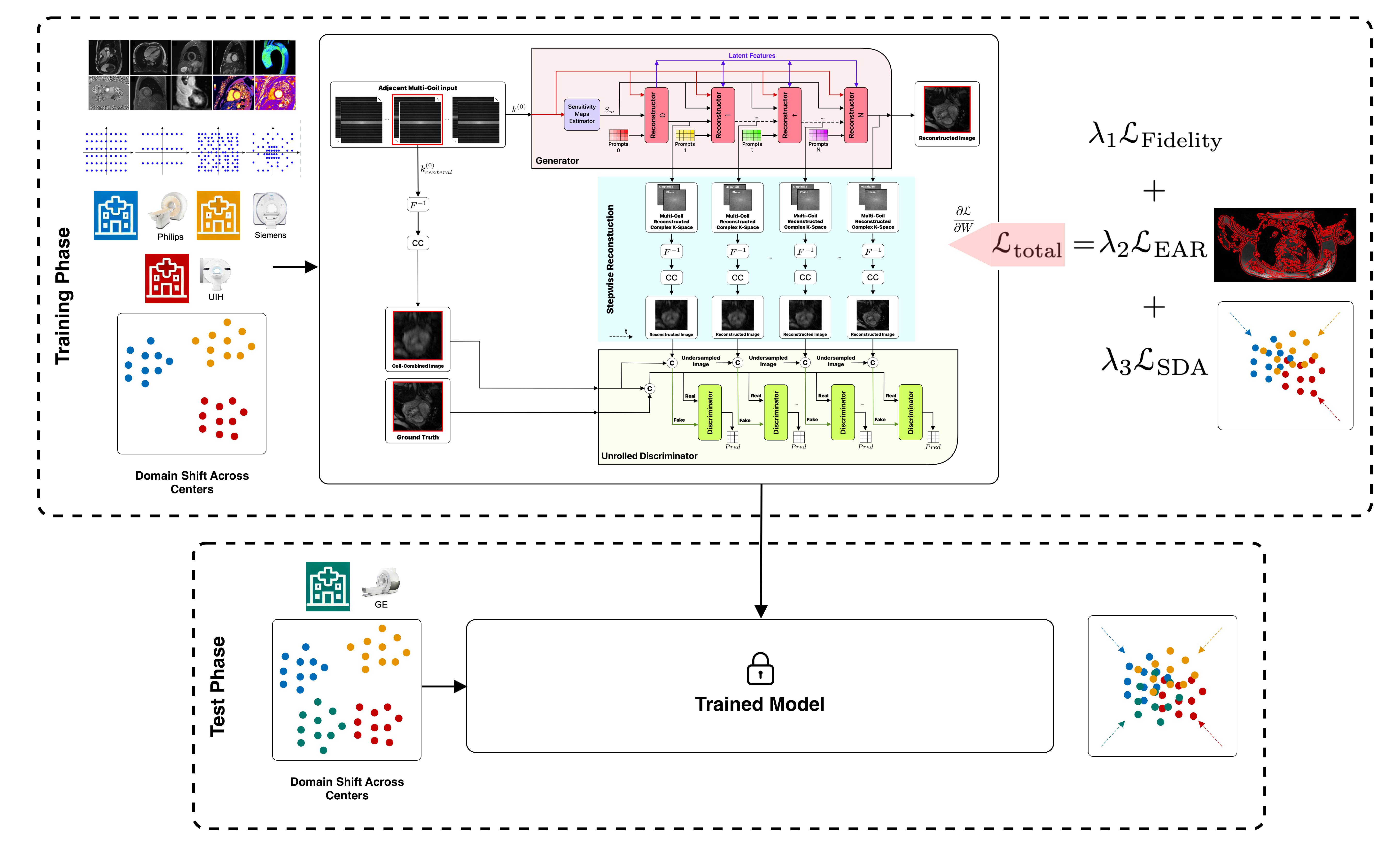}}
\caption{Overview of the GENRE-CMR method.}
\label{fig1}
\end{figure}

The total loss combines three components: a data fidelity loss 
($\mathcal{L}_{\text{Fidelity}}$), an edge-aware reconstruction loss 
($\mathcal{L}_{\text{EAR}}$), and a statistical distribution alignment loss 
($\mathcal{L}_{\text{SDA}}$), each weighted by hyperparameters 
$\lambda_1, \lambda_2,$ and $\lambda_3$, respectively. 
In our 2024 work \cite{anvari2024all}, we combined physical k-space consistency and 
SSIM losses. In 2025, we extend this formulation by introducing 
$\mathcal{L}_{\text{EAR}}$ for sharper edge preservation and 
$\mathcal{L}_{\text{SDA}}$ for explicit cross-domain alignment, 
resulting in the composite loss defined in Eq. \ref{eq1}.

\begin{equation}
\mathscr{L}_{\mathrm{total}} = \lambda_1 \mathscr{L}_{\mathrm{Fidelity}} + \lambda_2 \mathscr{L}_{\mathrm{EAR}} + \lambda_3 \mathscr{L}_{\mathrm{SDA}}
\label{eq1}
\end{equation}

The fidelity loss $\mathcal{L}_\text{Fidelity}$ corresponds to the total loss introduced in our previous work \cite{anvari2024all}, which combines image-domain and physical k-space domain consistency terms to ensure high-quality reconstruction. To better preserve fine anatomical structures and prevent blurring of high-frequency details in cardiac MR images, we propose EAR loss. This loss focuses on the local regions around the edges, where diagnostic features are most critical, isolating edge information and comparing ground-truth and reconstructed images only within these regions. First, we extract horizontal and vertical gradients from the ground-truth image $I_{\text{gt}}$ using $3 \times 3$ Sobel filters $S_x$ and $S_y$, producing gradient maps $G_x = S_x * I_{\text{gt}}$ and $G_y = S_y * I_{\text{gt}}$, where $*$ denotes convolution. The edge magnitude map is computed as

\begin{equation}
M = \sqrt{G_x^2 + G_y^2}
\label{eq2}
\end{equation}

To expand the influence area around edges, we convolve the edge map $M$ with a $5 \times 5$ averaging kernel $A$, resulting in a smoothed map $M_s = A * M$. After thresholding $M_s$ at $\tau=0$, a binary mask $B \in \{0,1\}^{H \times W}$ is generated as:

\begin{equation}
B_{i,j} = 
\begin{cases}
1 & \text{if } M_s(i,j) \geq \tau \\
0 & \text{otherwise}
\end{cases}
\label{eq3}
\end{equation}

This binary mask is then applied to both the reconstructed image $I_{\text{rec}}$ and the ground-truth $I_{\text{gt}}$, producing masked versions containing the regions of edges: $\tilde{I}_{\text{rec}} = B \odot I_{\text{rec}}$ and $\tilde{I}_{\text{gt}} = B \odot I_{\text{gt}}$, where $\odot$ denotes element-wise multiplication. Finally, the EAR loss is defined using the Structural Similarity Index (SSIM) loss between the masked images:

\begin{equation}
\mathcal{L}_{\text{EAR}} = 1 - \text{SSIM}(\tilde{I}_{\text{rec}}, \tilde{I}_{\text{gt}})
\label{eq4}
\end{equation}

\begin{figure}[!t]
\centerline{\includegraphics[width=0.85\textwidth]{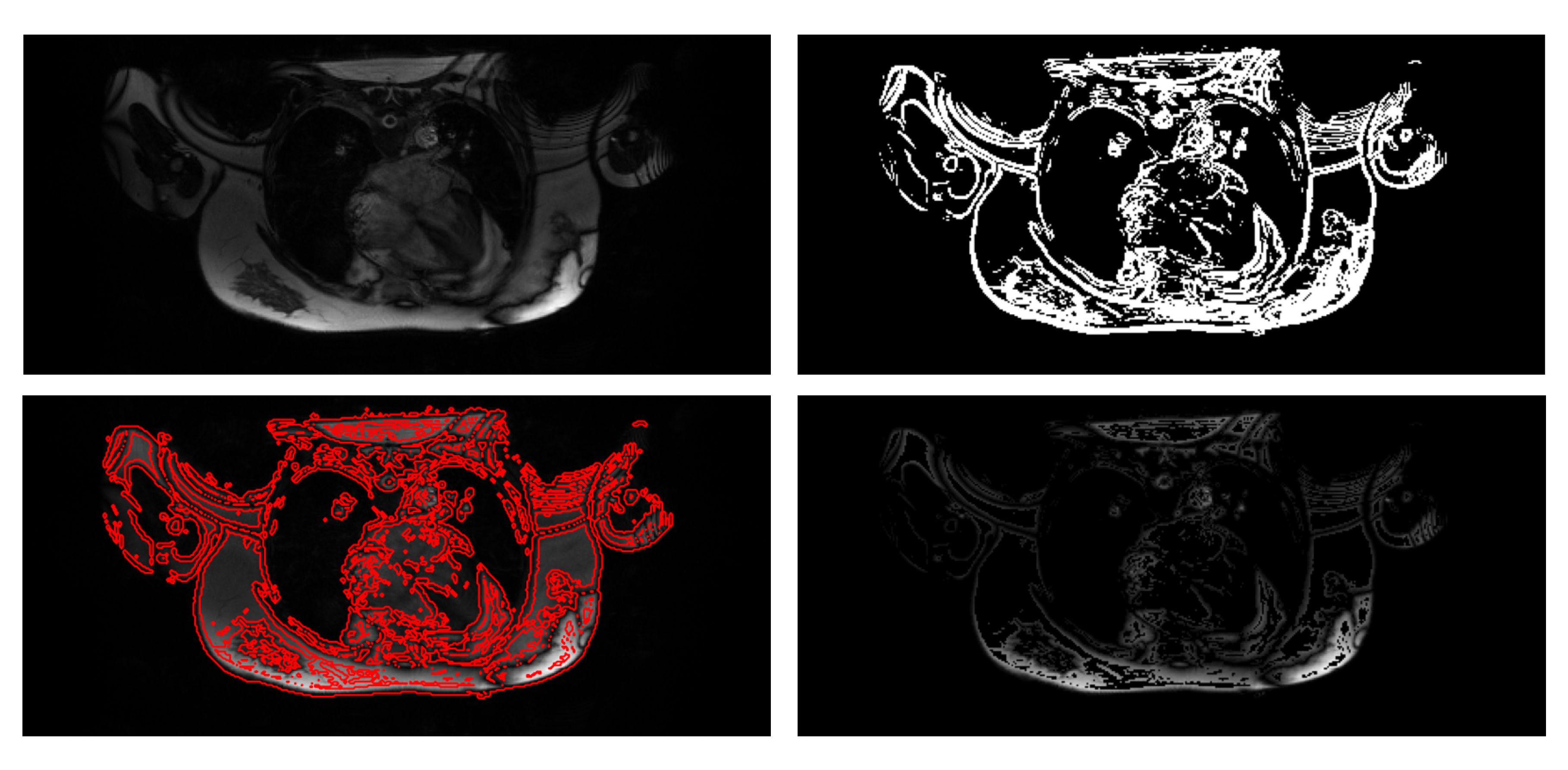}}
\caption{Illustration of the EAR Loss computation. Top-left: original/reconstructed image. Top-right: binary edge mask extracted from the image. Bottom-left: edge mask fused on the image. Bottom-right: image edges extracted using the mask.}
\label{fig2}
\end{figure}

This targeted design ensures that the model is penalized for structural degradation around edge regions, leading to sharper and more diagnostically valuable reconstructions. Illustration of the EAR Loss computations is shown in Figure \ref{fig2}.

The SDA loss encourages the intermediate feature representations of inputs from different domains to align in distribution. Suppose that the training data are drawn from five distinct distributions \( \mathcal{D}_1, \mathcal{D}_2, \dots, \mathcal{D}_5 \), and are processed sequentially so that each group of five consecutive samples \( \{x_t^{(1)}, x_t^{(2)}, \dots, x_t^{(5)}\} \) includes one data from each distribution. Each sample is passed through a residual deep unrolled network consisting of 16 reconstructors. For each sample \( x_t^{(i)} \), we extract feature vectors \( \mathbf{f}_t^{(i,l)} \in \mathbb{R}^d \) from the output of reconstructor \( l \in \{1, \dots, 16\} \). For each reconstructor, we compute a set \( \mathcal{F}_l = \{\mathbf{f}_t^{(1,l)}, \dots, \mathbf{f}_t^{(5,l)}\} \), assumed to be samples from domain-specific feature distributions.

Assuming these are multivariate Gaussian distributions \( \mathcal{N}(\mu_i^l, \Sigma_i^l) \) \cite{liang2025distribution}, we define SDA loss using symmetric KL-divergence:

\begin{equation}
\mathcal{L}_{\text{SDA}}^{(l)} = \frac{1}{10} \sum_{i<j} \left[ D_{\text{KL}}(\mathcal{N}(\mu_i^l, \Sigma_i^l) \| \mathcal{N}(\mu_j^l, \Sigma_j^l)) + D_{\text{KL}}(\mathcal{N}(\mu_j^l, \Sigma_j^l) \| \mathcal{N}(\mu_i^l, \Sigma_i^l)) \right]
\label{eq5}
\end{equation}

The total SDA loss is obtained by summing over all subnetwork layers:

\begin{equation}
\mathcal{L}_{\text{SDA}} = \sum_{l=1}^{16} \mathcal{L}_{\text{SDA}}^{(l)}
\label{eq6}
\end{equation}

To maintain temporal domain diversity, a sliding window mechanism is used during training: after the initial 5 samples (one per domain), for each new input sample, the SDA loss is calculated by comparing it with the 4 most recent samples from the training sequence. This enforces local alignment of domain-specific features and promotes domain-invariant representation learning.

\begin{algorithm*}
\caption{The Proposed GENRE-CMR Approach for CMRI Reconstruction}
\label{alg:genre}
\resizebox{1.2\textwidth}{!}{%
\begin{minipage}{1.2\textwidth}
\begin{algorithmic}[1]
\footnotesize 
\REQUIRE
\begin{tabular}[t]{@{}l@{}}
$k^{(0)}$: Adjacent multi-coil subsampled k-spaces \\
$k^{(G)}$: Ground-truth multi-coil k-space \\
$P^{(t)}$: Discriminative prompt at unroll step $t$ \\
$\mathscr{M}$: Subsampling mask \\
$\eta^{(t)}$: Learnable step size at unroll step $t$ \\
$\texttt{Label}_{real}$: Zero-filled matrix, $\ \texttt{Label}_{fake}$: One-filled matrix \\
$\lambda_1,\lambda_2,\lambda_3$: Loss weights (Fidelity, EAR, SDA) \\
$\texttt{SDAWindow}$: Sliding window size for domain alignment (e.g., $4$) \\
$\mathcal{D}$: Set of training domains (vendors / trajectories / contrasts)
\end{tabular}

\ENSURE
\begin{tabular}[t]{@{}l@{}}
$\theta_{Gen}$: Generator weights \\
$\theta_{Disc}$: Discriminator weights
\end{tabular}

\STATE $k^{(ACS)} \leftarrow \mathrm{Auto\_calibration\_signal}(k^{(0)})$; 
       $I^{(ACS)} \leftarrow \mathrm{FFT^{-1}}(k^{(ACS)})$
\STATE $S_m \leftarrow \mathrm{APUnet^{(SME)}}(I^{(ACS)})$; 
       $S_m' \leftarrow \mathrm{Conjugate\_Symmetry}(S_m)$ // Sensitivity estimation
\STATE Initialize residual feature $F^{(-1)} \leftarrow 0$, 
       feature banks $\{\mathcal{B}^{(l)}\}_{l=1}^{T}$ // Capacity \texttt{SDAWindow}
\STATE Let $T$ = number of unrolled steps (e.g., $16$)

\FOR{$\texttt{iteration}=1,2,\ldots$}
  \STATE $\mathcal{L}_{\text{Gen}} \leftarrow 0$
  \FOR{$t=0,1,\ldots,T-1$}
    \STATE $I_{MC} \leftarrow \mathrm{FFT^{-1}}(k^{(t)})$; 
           $I_{SC} \leftarrow S_m' \cdot I_{MC}$
\STATE $\hat{F}^{(t)} \leftarrow \mathrm{APUnet}^{(t)}(I_{SC}, P^{(t)})$; 
       $F^{(t)} \leftarrow \hat{F}^{(t)}+F^{(t-1)}$ // Residual unrolling
    \STATE $I'_{RF} \leftarrow \mathrm{Repeat\_interleaved}(F^{(t)})$; 
           $I_{SS} \leftarrow I'_{RF} \cdot S_m$; 
           $G_k \leftarrow \mathrm{FFT}(I_{SS})$
    \STATE $k^{(t+1)} \leftarrow k^{(t)} - \eta^{(t)} \mathscr{M}(k^{(t)}-k^{(0)}) + G_k$

    \STATE $\mathcal{L}^{(t)}_{\text{Phys}} \leftarrow \mathrm{MSE}\big(\mathrm{Mag}(k^{(t+1)}_{\text{central}}), \mathrm{Mag}(k^{(G)})\big) + \mathrm{MSE}\big(\Phi(k^{(t+1)}_{\text{central}}), \Phi(k^{(G)})\big)$
    \STATE $\mathcal{L}^{(t)}_{\text{SSIM}} \leftarrow \mathrm{SSIM\_Loss}\!\big(\mathrm{CC}(\mathrm{FFT^{-1}}(k^{(t+1)}_{\text{central}})), \mathrm{CC}(\mathrm{FFT^{-1}}(k^{(G)}))\big)$
    \STATE $\mathcal{L}^{(t)}_{\text{Fidelity}} \leftarrow \mathcal{L}^{(t)}_{\text{Phys}} + \mathcal{L}^{(t)}_{\text{SSIM}}$

    \STATE $I_{\text{rec}} \leftarrow \mathrm{CC}(\mathrm{FFT^{-1}}(k^{(t+1)}_{\text{central}}))$; 
           $I_{\text{gt}} \leftarrow \mathrm{CC}(\mathrm{FFT^{-1}}(k^{(G)}))$
    \STATE $M \leftarrow \sqrt{(\mathrm{Sobel}_x(I_{\text{gt}}))^2 + (\mathrm{Sobel}_y(I_{\text{gt}}))^2}$; 
           $M_s \leftarrow \mathrm{AvgPool}_{5\times5}(M)$; 
           $B \leftarrow \mathbb{1}[M_s \ge \tau]$
    \STATE $\tilde{I}_{\text{rec}} \leftarrow B \odot I_{\text{rec}}$; 
           $\tilde{I}_{\text{gt}} \leftarrow B \odot I_{\text{gt}}$
    \STATE $\mathcal{L}^{(t)}_{\text{EAR}} \leftarrow 1 - \mathrm{SSIM}(\tilde{I}_{\text{rec}}, \tilde{I}_{\text{gt}})$

    \STATE $z^{(t)} \leftarrow \mathrm{GlobalAvgPool}(F^{(t)})$; 
           $\mathrm{UpdateFeatureBank}(\mathcal{B}^{(t)}, z^{(t)}, \text{domain}(k^{(0)}))$
    \STATE $\mathcal{L}^{(t)}_{\text{SDA}} \leftarrow \mathrm{SymKL\_AcrossDomains}(\{\mathcal{B}^{(l)}\}_{l=1}^T,\texttt{SDAWindow})$

    \STATE $\mathcal{L}^{(t)}_{\text{step}} \leftarrow \lambda_1 \mathcal{L}^{(t)}_{\text{Fidelity}} + \lambda_2 \mathcal{L}^{(t)}_{\text{EAR}} + \lambda_3 \mathcal{L}^{(t)}_{\text{SDA}}$

    \STATE $\texttt{Input}_{real} \leftarrow \mathrm{Concat}(I_{\text{gt}}, \mathrm{CC}(\mathrm{FFT^{-1}}(k^{(0)}_{\text{central}})))$
    \STATE $\texttt{Input}_{fake} \leftarrow \mathrm{Concat}(I_{\text{rec}}, \mathrm{CC}(\mathrm{FFT^{-1}}(k^{(0)}_{\text{central}})))$
    \STATE $\texttt{Pred}_{real} \leftarrow \mathrm{Discriminator}(\texttt{Input}_{real})$; 
           $\texttt{Pred}_{fake} \leftarrow \mathrm{Discriminator}(\texttt{Input}_{fake})$
    \STATE $\mathcal{L}_{real} \leftarrow \mathrm{BCE\_Loss}(\texttt{Label}_{real}, \texttt{Pred}_{real})$; 
           $\mathcal{L}_{fake} \leftarrow \mathrm{BCE\_Loss}(\texttt{Label}_{fake}, \texttt{Pred}_{fake})$
    \STATE $\mathcal{L}_{Disc} \leftarrow \tfrac{1}{2}(\mathcal{L}_{real}+\mathcal{L}_{fake})$; 
           $\theta_{Disc} \leftarrow \mathrm{AdamW}(\mathcal{L}_{Disc}, \theta_{Disc})$

    \STATE $\mathcal{L}_{\text{Gen}} \leftarrow \mathcal{L}_{\text{Gen}} + \mathcal{L}^{(t)}_{\text{step}}$
  \ENDFOR

  \STATE $\texttt{Input}_{fake} \leftarrow \mathrm{Concat}(\mathrm{CC}(\mathrm{FFT^{-1}}(k^{(T)}_{\text{central}})), \mathrm{CC}(\mathrm{FFT^{-1}}(k^{(0)}_{\text{central}})))$
  \STATE $\texttt{Pred}_{fake} \leftarrow \mathrm{Discriminator}(\texttt{Input}_{fake})$
  \STATE $\mathcal{L}_{\text{GAN-Gen}} \leftarrow \mathrm{BCE\_Loss}(\texttt{Label}_{real}, \texttt{Pred}_{fake})$
  \STATE $\mathcal{L}_{\text{Total-Gen}} \leftarrow \mathcal{L}_{\text{Gen}} + \mathcal{L}_{\text{GAN-Gen}}$
  \STATE $(\lambda_1,\lambda_2,\lambda_3) \leftarrow \mathrm{UpdateLossWeightsByCoV}(\{\mathcal{L}^{(t)}_{\text{Fidelity}}\},\{\mathcal{L}^{(t)}_{\text{EAR}}\},\{\mathcal{L}^{(t)}_{\text{SDA}}\})$
  \STATE $\theta_{Gen}, \theta_{SME} \leftarrow \mathrm{AdamW}(\mathcal{L}_{\text{Total-Gen}}, \theta_{Gen}, \theta_{SME})$
\ENDFOR

\RETURN $\theta_{Gen}$ // Final generator weights
\end{algorithmic}
\end{minipage}
}
\end{algorithm*}

\subsection{Implementation Details}

\begin{figure}[!t]
\centerline{\includegraphics[width=0.7\textwidth]{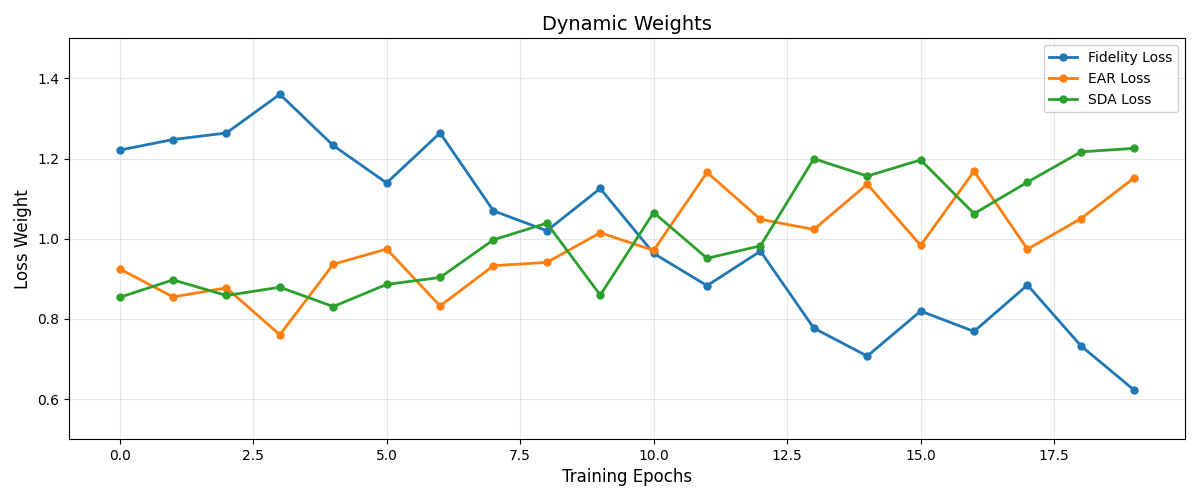}}
\caption{Dynamic weighting of loss components during training. The plot illustrates how each loss weight changes over time, enabling adaptive emphasis on different loss terms throughout the optimization process.}
\label{fig3}
\end{figure}

In our approach, the generator and discriminator were optimized with AdamW (learning rate 0.002, weight decay 0.1, gradient clipping 0.1, step scheduler with step size 11 and gamma 0.1). The generator uses 16 reconstructor modules, 16 auto-calibration lines, and an adjacent k-space length of 5. We used acceleration factors of 8, 16, and 24 with k-t uniform, k-t Gaussian, and k-t radial trajectories. Curriculum learning \cite{bengio2009curriculum} was applied by starting with lower acceleration factors, training for 20 epochs with batch size 1. Model performance was evaluated using SSIM, PSNR, and NMSE. To balance loss terms, we adopted Coefficient of Variation (CoV) weighting \cite{9423366}, which dynamically adjusts each loss weight ($\lambda_1$, $\lambda_2$, $\lambda_3$) based on the ratio of standard deviation to mean, giving higher priority to more variable losses. The evolution of these weights is shown in Figure~\ref{fig3}.

\section{Experimental Results and Discussion}
\begin{figure}[!t]
\centerline{\includegraphics[width=0.95\textwidth]{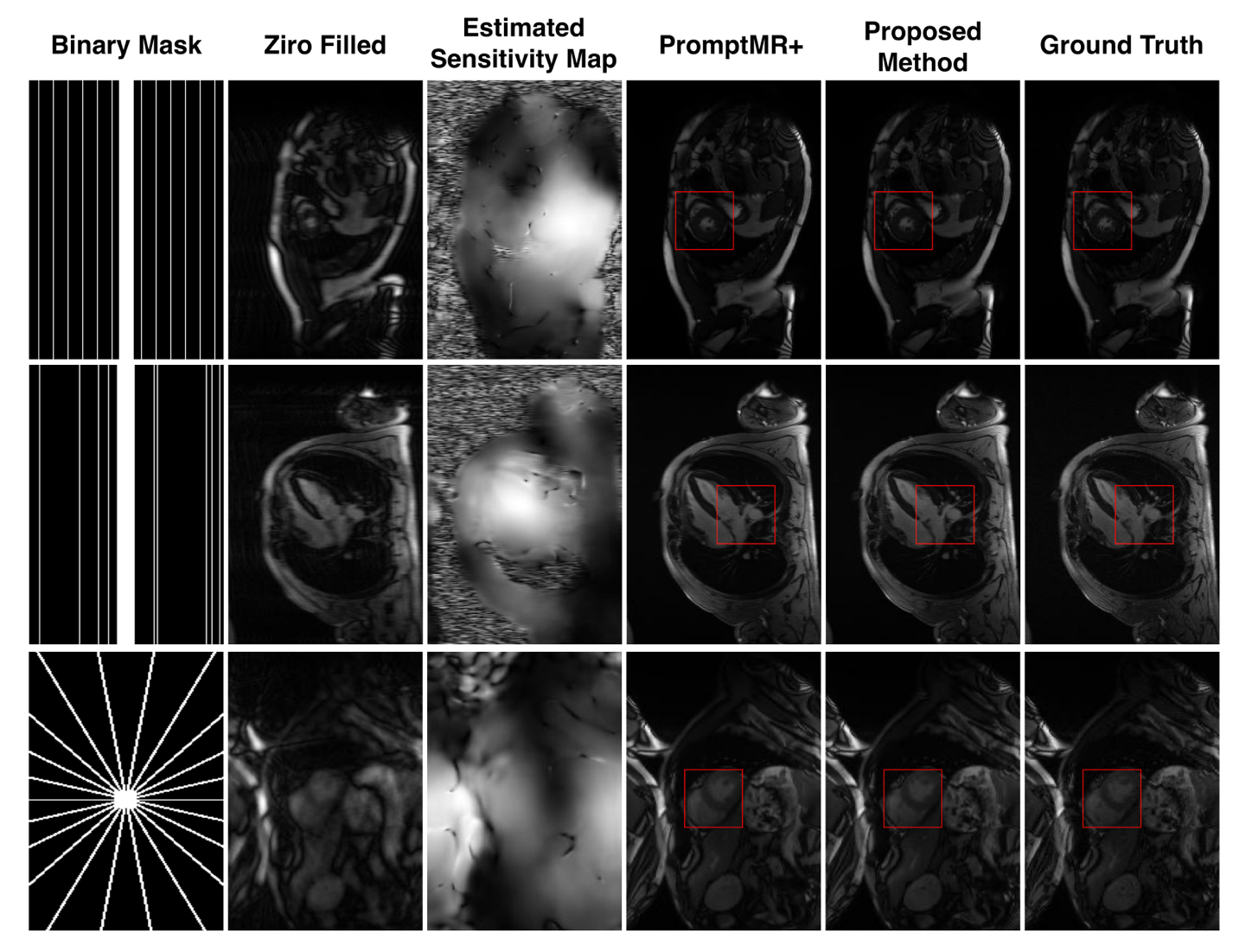}}
\caption{Qualitative comparison of reconstruction results under three different sampling trajectories at an acceleration factor of 16. }
\label{fig4}
\end{figure}

To evaluate the performance of our proposed method, we conducted extensive experiments using both seen (training) and unseen distributions. The quantitative results are summarized in Table \ref{tab1}, and the effectiveness of individual architectural components is examined through ablation studies in Table \ref{tab2}. A qualitative comparison of reconstructed images is also provided in Figure \ref{fig4} under three distinct sampling trajectories at an acceleration factor of 16.

\begin{table*}[!htp]
\centering
\caption{State-of-the-art MRI reconstruction approaches versus the proposed method. Metrics include SSIM, PSNR, and NMSE. The best values are shown in bold.}
\label{tab1}
\renewcommand{\arraystretch}{1.5}
\setlength{\tabcolsep}{10pt} 
\resizebox{0.8\textwidth}{!}{
\begin{tabular}{l|ccc|ccc}
\toprule
\multirow{2}{*}{\textbf{Method}} & \multicolumn{3}{c|}{\textbf{Training Distributions}} & \multicolumn{3}{c}{\textbf{Unseen Distributions}} \\
\cmidrule(lr){2-4} \cmidrule(lr){5-7}
 & SSIM & PSNR & NMSE & SSIM & PSNR & NMSE \\
\midrule
PromptMR \cite{xin2023fill}        & 0.9685 & 41.80 & 0.0129 & 0.9450 & 37.85 & 0.0198 \\
SR-GAN \cite{anvari2024all}         & 0.9702 & 42.05 & 0.0120 & 0.9473 & 38.01 & 0.0191 \\
PromptMR+ \cite{xin2024rethinking}       & 0.9728 & 42.40 & 0.0115 & 0.9498 & 38.22 & 0.0187 \\
\textbf{GENRE-CMR} & \textbf{0.9743} & \textbf{42.64} & \textbf{0.0111} & \textbf{0.9552} & \textbf{38.90} & \textbf{0.0160} \\
\bottomrule
\end{tabular}
}
\end{table*}

Table \ref{tab1} presents a comparative evaluation of our method, GENRE-CMR, against several state-of-the-art approaches, including PromptMR \cite{xin2023fill}, SR-GAN \cite{anvari2024all} and PromptMR+ \cite{xin2024rethinking}. Our model achieves the highest scores across all three metrics, SSIM, PSNR, and NMSE, in both training and unseen distributions. Specifically, GENRE-CMR attains a PSNR of 42.64 dB and SSIM of 0.9743 on training distributions, and maintains strong generalization with 38.90 dB PSNR and 0.9552 SSIM on unseen data. These results indicate that GENRE-CMR learns effective representations from the training data while maintaining robust performance on unseen data distributions. The consistent reduction in NMSE further demonstrates the model’s ability to preserve fine image details and suppress reconstruction errors.

\begin{table*}[!htp]
\centering
\caption{Ablation study results demonstrating the impact of different components in the proposed method. Evaluation metrics include SSIM, PSNR, and NMSE, with the best scores highlighted in bold.}
\label{tab2}
\renewcommand{\arraystretch}{1.5}
\setlength{\tabcolsep}{10pt} 
\resizebox{0.6\textwidth}{!}{
\begin{tabular}{l|ccc}
\toprule
\multirow{2}{*}{\textbf{Ablation}} & \multicolumn{3}{c}{\textbf{Unseen Distributions}} \\
\cmidrule(lr){2-4}
 & SSIM & PSNR & NMSE \\
\midrule
Baseline & 0.9473 & 38.01 & 0.0191 \\
Without SDA Loss & 0.9500 & 38.25 & 0.0183 \\
Without EAR Loss & 0.9515 & 38.43 & 0.0178 \\
Without Residual Connections & 0.9523 & 38.21 & 0.0171 \\
\textbf{The Proposed Method} & \textbf{0.9552} & \textbf{38.90} & \textbf{0.0160} \\
\bottomrule
\end{tabular}
}
\end{table*}

As illustrated in Figure \ref{fig4}, the proposed method provides visually superior reconstructions compared to existing techniques under various undersampling trajectories. The reconstructed images by GENRE-CMR show sharper anatomical boundaries and reduced aliasing artifacts, particularly in challenging regions such as myocardial borders and small vessels. This visual fidelity underscores the benefit of integrating residual unrolling with edge-aware reconstruction.

To assess the contribution of individual components in our architecture, we performed a systematic ablation study. Table \ref{tab2} reports the performance on unseen distributions when certain modules are removed from the full pipeline. Removing the SDA loss resulted in a noticeable performance drop, confirming its role in improving cross-distribution generalization. Similarly, eliminating the EAR loss led to reduced SSIM and PSNR, indicating its effectiveness in enhancing edge preservation during reconstruction. Disabling residual connections between cascaded subnetworks also caused a decline in performance, confirming that feature propagation across depths is crucial for effective hierarchical representation learning. In the full configuration, the proposed method outperforms all ablated variants, achieving the highest SSIM (0.9552), PSNR (38.90), and the lowest NMSE (0.0160). This demonstrates the synergistic benefit of combining residual connections, EAR loss, and SDA loss in our unrolled reconstruction framework. Compared to our 2024 all-in-one model \cite{anvari2024all}, the proposed 
GENRE-CMR consistently outperforms it under both in-distribution and 
out-of-distribution evaluation. Ablation studies confirm that 
$\mathcal{L}_{\text{EAR}}$ and $\mathcal{L}_{\text{SDA}}$, absent in the 2024 version, 
are responsible for substantial gains in edge fidelity and domain robustness.

\section{Conclusion}
In this work, we proposed a generalizable deep learning framework for CMR reconstruction, built upon a GAN architecture. The generator is driven by a residual deep unrolled architecture that mimics iterative optimization steps, incorporating compressed sensing concepts to effectively address the underlying inverse problem. To enhance the preservation of clinically relevant anatomical structures, we introduced the EAR loss, which explicitly promotes sharper boundary reconstruction and reduces common blurring artifacts. We integrated the SDA loss to improve robustness across diverse distributions to address domain shift challenges that arise due to differences in imaging centers and devices, image contrast, sampling patterns, anatomical variability, and acquisition protocols. Comprehensive experiments and ablation studies on the CMRxRecon 2025 dataset demonstrated that our method consistently outperforms state-of-the-art approaches. 

While effective, the framework can be computationally demanding during training, requiring powerful GPUs to achieve optimal performance. This is a common challenge in advanced deep learning methods, and future work may explore more efficient architectures or training strategies to reduce resource requirements. As an important future direction, we plan to conduct clinical evaluations with expert radiologists to assess diagnostic accuracy and real-world applicability, moving closer to clinical integration.

\bibliographystyle{splncs04}

\end{document}